\documentclass[10pt]{paper}
\usepackage{amsmath, amsthm, amssymb, }
\usepackage{float,epsfig}
\usepackage{color}
\usepackage{graphicx}
\usepackage{mathtools}
\usepackage{subcaption}
\RequirePackage{graphicx,epstopdf,adjustbox}

\begin{document}

\newtheorem{theorem}{\bf Theorem}[section]
\newtheorem{proposition}[theorem]{\bf Proposition}
\newtheorem{definition}[theorem]{\bf Definition}
\newtheorem{corollary}[theorem]{\bf Corollary}
\newtheorem{remark}[theorem]{\bf Remark}
\newtheorem{lemma}[theorem]{\bf Lemma}
\newcommand{\nrm}[1]{|\!|\!| {#1} |\!|\!|}

\newcommand{\ba}{\begin{array}}
\newcommand{\ea}{\end{array}}
\newcommand{\von}{\vskip 1ex}
\newcommand{\vone}{\vskip 2ex}
\newcommand{\vtwo}{\vskip 4ex}
\newcommand{\dm}[1]{ {\displaystyle{#1} } }

\newcommand{\be}{\begin{equation}}
\newcommand{\ee}{\end{equation}}
\newcommand{\beano}{\begin{eqnarray*}}
\newcommand{\eeano}{\end{eqnarray*}}
\newcommand{\inp}[2]{\langle {#1} ,\,{#2} \rangle}
\def\bmatrix#1{\left[ \begin{matrix} #1 \end{matrix} \right]}
\def \noin{\noindent}
\newcommand{\evenindex}{\Pi_e}



\def \R{{\mathbb R}}
\def \C{{\mathbb C}}
\def \F{{\mathbb F}}
\def \J{{\mathbb J}}
\def \L{\mathcal{L}}

\def \T{{\mathbb T}}
\def \Pb{\mathrm{P}}
\def \N{{\mathbb N}}
\def \Ib{\mathrm{I}}
\def \Ls{{\Lambda}_{m-1}}
\def \Gb{\mathrm{G}}
\def \Hb{\mathrm{H}}
\def \Lam{{\Lambda}}
\def \Qb{\mathrm{Q}}
\def \Rb{\mathrm{R}}
\def \Mb{\mathrm{M}}
\def \norm{\nrm{\cdot}\equiv \nrm{\cdot}}

\def \P{{\mathbb{P}}_m(\C^{n\times n})}
\def \A{{{\mathbb P}_1(\C^{n\times n})}}
\def \H{{\mathbb H}}
\def \L{{\mathbb L}}
\def \G{{\F_{\tt{H}}}}
\def \S{\mathbb{S}}
\def \sigmin{\sigma_{\min}}
\def \elam{\Lambda_{\epsilon}}
\def \slam{\Lambda^{\S}_{\epsilon}}
\def \Ib{\mathrm{I}}
\def \Tb{\mathrm{T}}
\def \d{{\delta}}

\def \Lb{\mathrm{L}}
\def \N{{\mathbb N}}
\def \Ls{{\Lambda}_{m-1}}
\def \Gb{\mathrm{G}}
\def \Hb{\mathrm{H}}
\def \Delta{\triangle}
\def \Rar{\Rightarrow}
\def \p{{\mathsf{p}(\lam; v)}}

\def \D{{\mathbb D}}

\def \tr{\mathrm{Tr}}
\def \cond{\mathrm{cond}}
\def \lam{\lambda}
\def \sig{\sigma}
\def \sign{\mathrm{sign}}

\def \ep{\epsilon}
\def \diag{\mathrm{diag}}
\def \rev{\mathrm{rev}}
\def \vec{\mathrm{vec}}

\def \herm{\mathsf{Herm}}
\def \sym{\mathsf{sym}}
\def \odd{\mathsf{sym}}
\def \en{\mathrm{even}}
\def \rank{\mathrm{rank}}
\def \pf{{\bf Proof: }}
\def \dist{\mathrm{dist}}
\def \rar{\rightarrow}

\def \rank{\mathrm{rank}}
\def \pf{{\bf Proof: }}
\def \dist{\mathrm{dist}}
\def \Re{\mathsf{Re}}
\def \Im{\mathsf{Im}}
\def \re{\mathsf{re}}
\def \im{\mathsf{im}}
\def \sp{\mathsf{Spec}}

\def \sym{\mathsf{sym}}
\def \sksym{\mathsf{skew\mbox{-}sym}}
\def \odd{\mathrm{odd}}
\def \even{\mathrm{even}}
\def \herm{\mathsf{Herm}}
\def \skherm{\mathsf{skew\mbox{-}Herm}}
\def \str{\mathrm{ Struct}}
\def \eproof{$\blacksquare$}
\def \proof{\noin\pf}

\def \bS{{\bf S}}
\def \calSC{{\mathcal{SC}}}
\def \E{{\mathcal E}}
\def \X{{\mathcal X}}
\def \cH{\mathcal{H}}
\def \cJ{\mathcal{J}}
\def \tr{\mathrm{Tr}}
\def \range{\mathrm{Range}}
\def \adj{\star}
\def \diag{\mbox{diag}}
\def \rank{\mbox{rank}}

\def \pal{\mathrm{palindromic}}
\def \palpen{\mathrm{palindromic~~ pencil}}
\def \palpoly{\mathrm{palindromic~~ polynomial}}
\def \odd{\mathrm{odd}}
\def \even{\mathrm{even}}

\title{Why lockdown : On the spread of SARS-CoV-2 in India, a network approach}
\author{Pradumn Kumar Pandey\thanks{Department of Computer Science and Engineering, IIT Roorkee, India Email: pradumn.pandey@cs.iitr.ac.in}\, and  Bibhas Adhikari\thanks{Corresponding author, Department of Mathematics and Center for Theoretical Studies, IIT Kharagpur, India, E-mail: bibhas@maths.iitkgp.ac.in }}

\date{}

\maketitle
\thispagestyle{empty}
\noindent{\bf Abstract:} We analyze the time series data of number of districts or cities in India that are affected by COVID-19 from March 01, 2020 to April 17, 2020. We study the data in the framework of time series network data. The networks are defined by using the geodesic distances of the districts or cities specified by the latitude and longitude coordinates. We particularly restrict our analysis to all but districts in the north-eastern part of India. Unlike recent studies on the projection of the number of people infected with SARS-CoV-2 in the near future, in this note, the emphasis is on understanding the dynamics of the spread of the virus across the districts of India.

We perform spectral and structural analysis of the model networks by considering several measures, notably the spectral radius, the algebraic connectivity, the average clustering coefficient, the average path length and the structure of the communities. Furthermore, we study the overall expansion properties given by the number of districts or cities before and after lockdown. These studies show that lockdown has a significant impact on the spread of SARS-CoV-2 in districts or cities over long distances. However, this impact is only observed after approximately two weeks of lockdown. We speculate that this happened due to the insufficient number of tests for COVID-19 before the lockdown which could not stop the movement of people infected with the virus but not detected, over long distances. \\


\section{Introduction}

In this note, we study the dynamics of the spread of the SARS-CoV-2 in India before and after the announcement of a complete nationwide lockdown by the Indian government from March 25, 2020. The lockdown was announced by the Indian government for a period of 21 days initially as a preventive measure to stop the spread of the SARS-CoV-2. However, the lockdown was extended until May 03, 2020. Prior to that, on March 22, 2020 the Indian government decided to completely lock 82 districts in 22 states, and on March 23, 2020 the union and state governments announced the lockdown of 75 districts where corona positive cases have been reported. There are 28 states and 8 union territories in India. The states and territories of the union are subdivided into districts. There are a total of 718 districts in India. The first confirmed case of corona infection was reported on January 30, 2020 in the district of Thrissur in the state of Kerala which is a state of southern India. The number of people infected with SARS-CoV-2 was around 500 when the national lockdown was declared on March 25, 2020 in the first phase.

The data analyzed in this article comes from How India Lives (HIL), which is a definitive repository of public data in India \cite{hil}. They triangulate data from Indian government and nodal agencies, including the National Disaster Management Authority, the Ministry of Health and Welfare and the respective State Departments. We adapt the techniques of network data analysis for investigation of the spread of SARS-CoV-2 across the districts in India. We restrict our attention to all but the districts in the north-eastern part of India. 
This is done to reduce the density of the model networks, because the density depends on the distances between districts or cities and most districts in northeast India are far from the rest of the part of the country.


We first build a time series network data from the time series data of the districts in which at least one person infected with the coronavirus lives, on each date since March 01, 2020. Thus, the networks that we study in this note have sets of vertices which are districts or cities. According to the record, no positive case of infection was reported during the period from February 4 to March 1, 2020. There were only 3 cases of infection from January 30 to February 3, 2020 in the first place, and all of these cases are from the Kerala district. Another was detected with the virus in the city of Hyderabad, in the state of Telengana on March 3, then on March 4, several cases were reported in Delhi and in the state of Uttar Pradesh. Since then, on each date, several people from different districts are detected with the COVID-19, and thus a time series of an increasing number of districts is obtained. The edges in the proposed network data $N(t)$ are defined based on the geodesic distance between the districts/cities. Here we mention that we use the words graph and network interchangeably in this note. 

Let $A_1$ and $A_2$ be two districts or cities specified by the coordinates $(\phi_1, \lambda_1)$ and $(\phi_2, \lambda_2)$ respectively, where $\phi_i$ denotes the latitude and $\lambda_i$ denotes the longitude of $A_i, i=1,2$. We use the MATLAB inbuit function ``distance" to compute the great circle arcs connecting these pair of points on the surface of a sphere. For more details refer to \cite{dist}. Indeed, the distance $d$ between two points specified by their longitude and latitude is calculated using `haversine' formula which calculates shortest distance on the surface assuming it is smooth, there is no hills and crests. The distance $d$ is the great-circle distance between the two points $A_1$ and $A_2$ can be computed as described in  \cite{web}. 


Let $V (t)$ be the vertex set on day $t \geq 0 $ which is made up of all the districts in which at least one person is infected. Then $i, j\in V(t)$ are linked by an edge in $N(t)$ if the geodesic distance between $i$ and $j$ is less or equal to a value $d(t),$ which is the minimum value such that the edges form a connected network (see Section \ref{sec:2}). 
Here, we mention that two neighboring districts of $V(t)$ do not always need to be connected by an edge, on the other hand, the districts of $V(t)$ which belong to the same state are not necessarily connected by an edge. We call $d(t)$ as the {\textit{connectivity parameter}} of the network $N(t).$ 

An immediate observation that follows from the definition of the proposed time series networks is that when no restrictions are placed on travelers across the country, a potential carrier can travel a long distance before being detected with COVID-19. Therefore, such a spreader can infect other people living at a great distance. The important issue here is to analyze the value of $d(t)$ when $t $ is a date during lockdown. More importantly, how are these values distributed over the entire period from January 30, 2020? Some other relevant questions: What are the changes in the structural properties of these networks before and after lockdown? Does the lockdown prevent the virus from spreading across districts? How does the community structure of the network evolve over time?

First we plot the number of districts over time where at least one person is infected with the virus. A curve fitting technique finds that the growth is cubic before the lockdown, and cumulative number of infected districts or cities follows the growth of hyperbolic tangent during the entire period. Here we mention that the hyperbolic tangent formula for number of infected people in a network under SIS model is an well known fact. Then after calculating the value of the connectivity parameter $d(t)$ of $N(t)$ it shows that it gradually decreases from March 3, 2020. From these studies, we comprehend that sudden lifting of lockdown can speed up the spread to a large extent. 


The data of \textit{maximum degree} and \textit{medium degree} of $N(t)$ show that there is a change-point towards the date of lockdown. The maximum degree is a linear increase as it continues to infect neighboring districts until approximately 2 weeks after lockdown. On the other hand, the average number of newly infected neighboring districts of all the districts, given by the average degree of the networks, fluctuates up to two weeks after lockdown, then stabilizes. Next, we consider two spectral properties, \textit{spectral radius} ($\rho(t)$) and \textit{algebraic connectivity} ($\lam(t)$) of networks which provide important information about the structural properties of the networks, and play important role into the analysis of diffusion phenomena on networks. We observe from our simulation that $\rho (t) $ continues to increase up to two weeks after lockdown and then stabilizes. While the calculation of $\lam(t)$ shows that it gradually decreases until lockdown, which means that infected cases have occurred in distant places, but after lockdown, $\lam (t) $ behaves as an increasing function, which means that the infection is spread evenly across the districts. Further, we investigate the average clustering coefficients, number of triangles, diameter, average path length of the networks $N(t)$ and these data give a similar observation as discussed above.

Finally, we perform the \textit{community} analysis of $N(t).$ Note that a collection of communities on a network is given by a collection of dense sub-graphs so that there are only a few edges that connect separate communities. We observe a strong community structure of the $N(t)$ as measured by the modularity values. The number of communities behaves as a step function,and gradually increases before the lockdown and, after two weeks of lockdown, the step values gradually decrease, which means that the newly affected districts belong to the same community or different communities merge. This implies a homogeneous growth of the new affected districts across India, which is confirmed by the analysis of the size of the communities.

Thus, a common phenomenon that can be observed after examining the various measures described above is that a significant effect of the lockdown on these measures does not occur until after two weeks. This indicates that the spread has occurred since people with COVID-19 were not identified until the lockdown. The current growth of the spread could have been prevented if enough people with recent travel history had been tested before the lockdown. Our study also reveals that a national lockdown can reduce the spread of the virus over very long distances, but it helps only a little to stop the spread to neighboring districts. This can happen for several reasons. An obvious reason is that movement between districts without being detected as a carrier of the coronavirus. The spread can be prevented by establishing testing facilities at the district boundaries, and limited people should be allowed to travel a long distance once the test is negative, which appears to be a comprehensive exercise.

\section{Analysis of the spread of SARS-CoV-2 across districts in India}\label{sec:2}

In this section we make certain observation of the time series data of districts with at least one SARS-CoV-2 infected person during the period of January 30, 2020 - April 18, 2020. Time series networks $N(t)$ are generated based on this data which provide a platform for collective analysis of the data. In $N(t),$ the vertex set denoted by $V(t)$ represents a set of districts with at least one person is living in that district or city with COVID-19. Two vertices are linked by an edge if the geodesic distance of the vertices is less than or equal to a value $d(t)$, called connectivity parameter of $N(t),$  and this value is the minimum distance among the vertices for which the network $N(t)$ becomes connected. 

Thus the adjacency matrix $A(t)=[a_{ij}(t)]$ of $N(t)$ is defined as follows. Let $D=[d_{ij}]$ be the distance matrix for the vertex set, where $d_{ij}$ represents the distance between the districts $i$ and $j.$ Then $a_{ij}(t)=1, i\neq j$ if $d_{ij}\leq d(t),$ otherwise $a_{ij}(t)=0$. The eigenvalues of $A(t)$ are real since $A(t)$ is a symmetric matrix, and hence they can be ordered. The Laplacian matrix $L(t)=[l_{ij}(t)]$ associated with $N(t)$ is given by $$l_{ij}(t)=\sum_{j}a_{ij}(t) - a_{ij}(t).$$ $L(t)$ is a symmetric positive semi-definite matrix, and hence its eigenvalues are nonnegative and can be ordered. The smallest eigenvalue of $L(t)$ is $0.$ 

Below we study structural and statistical properties of $N(t)$ in order to investigate the effect of lockdown on the spread of SARS-CoV-2 across districts in India.

\subsection{Overall expansion} 

\begin{figure}[ht]
\begin{subfigure}{.5\textwidth}
  \centering
  \includegraphics[width=1\linewidth]{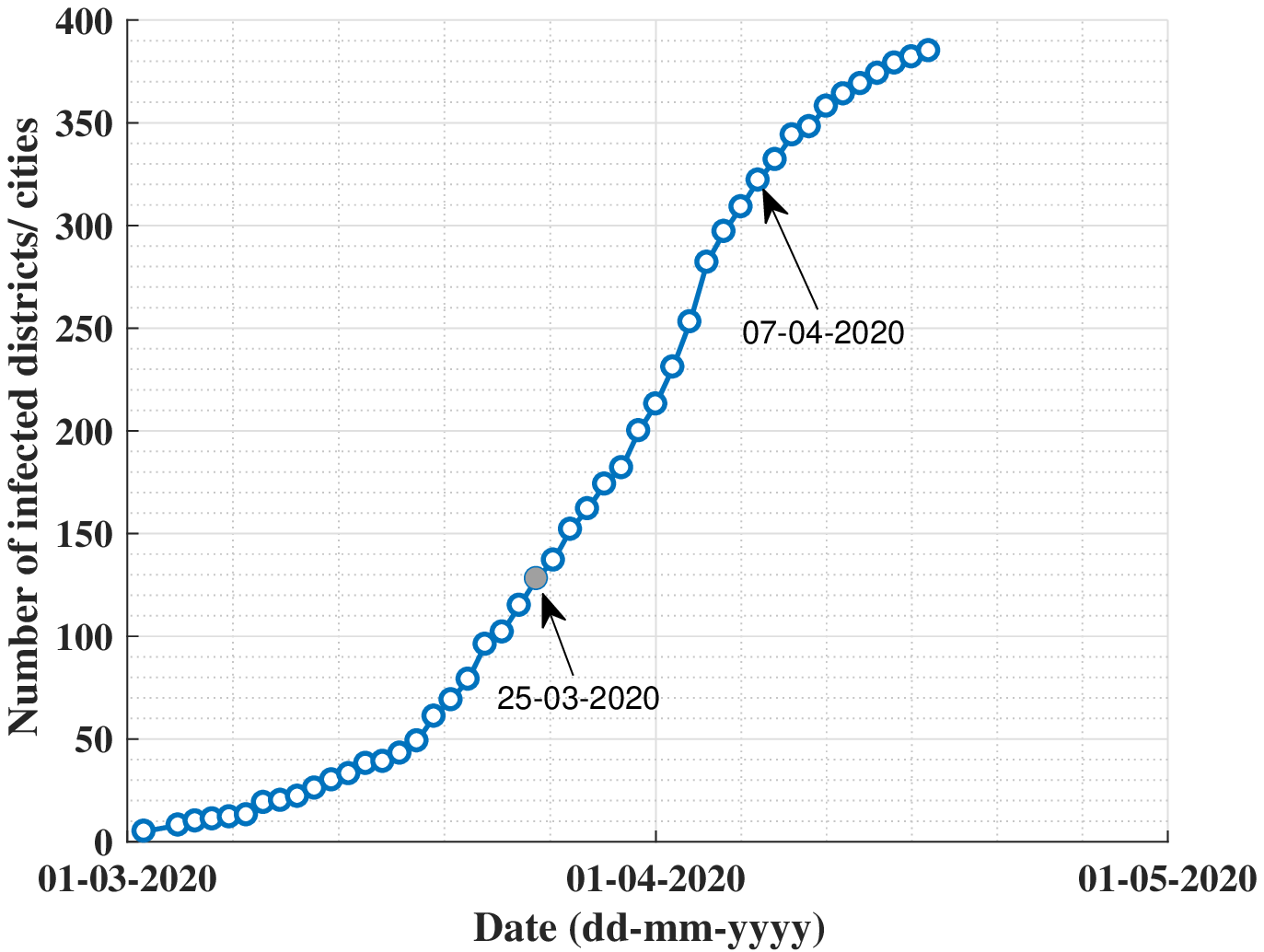}  
  \caption{}
  \label{fig1}
\end{subfigure}
\begin{subfigure}{.5\textwidth}
  \centering
  \includegraphics[width=1\linewidth]{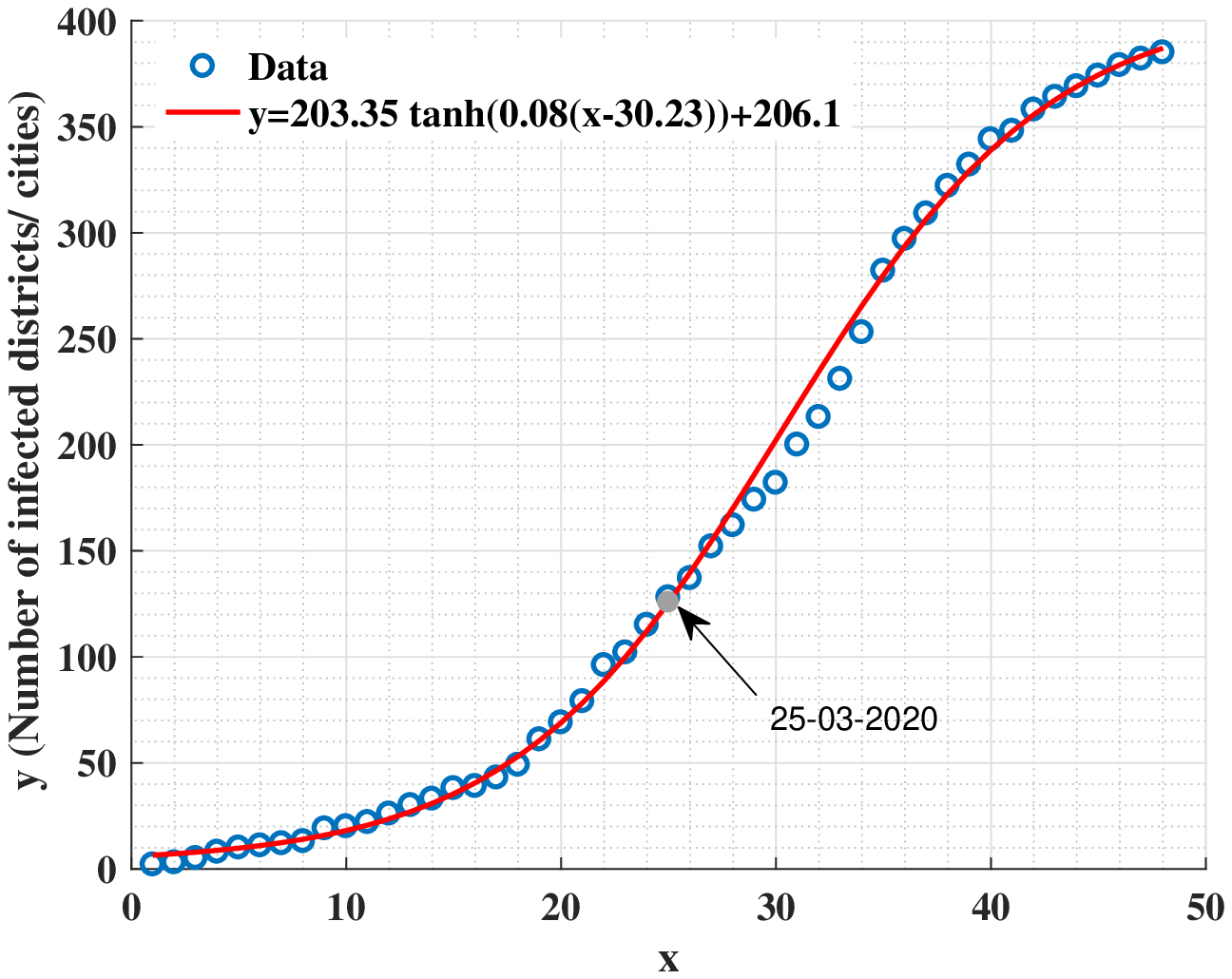}  
  \caption{}
  \label{fig16}
\end{subfigure}
\caption{(a) Cumulative number of infected districts or cities. (b) Data fitting using $\tan$ hyperbolic function.}
\label{fig:fig}
\end{figure}

\begin{figure}[ht]
\begin{subfigure}{.5\textwidth}
  \centering
 \includegraphics[width=1\linewidth]{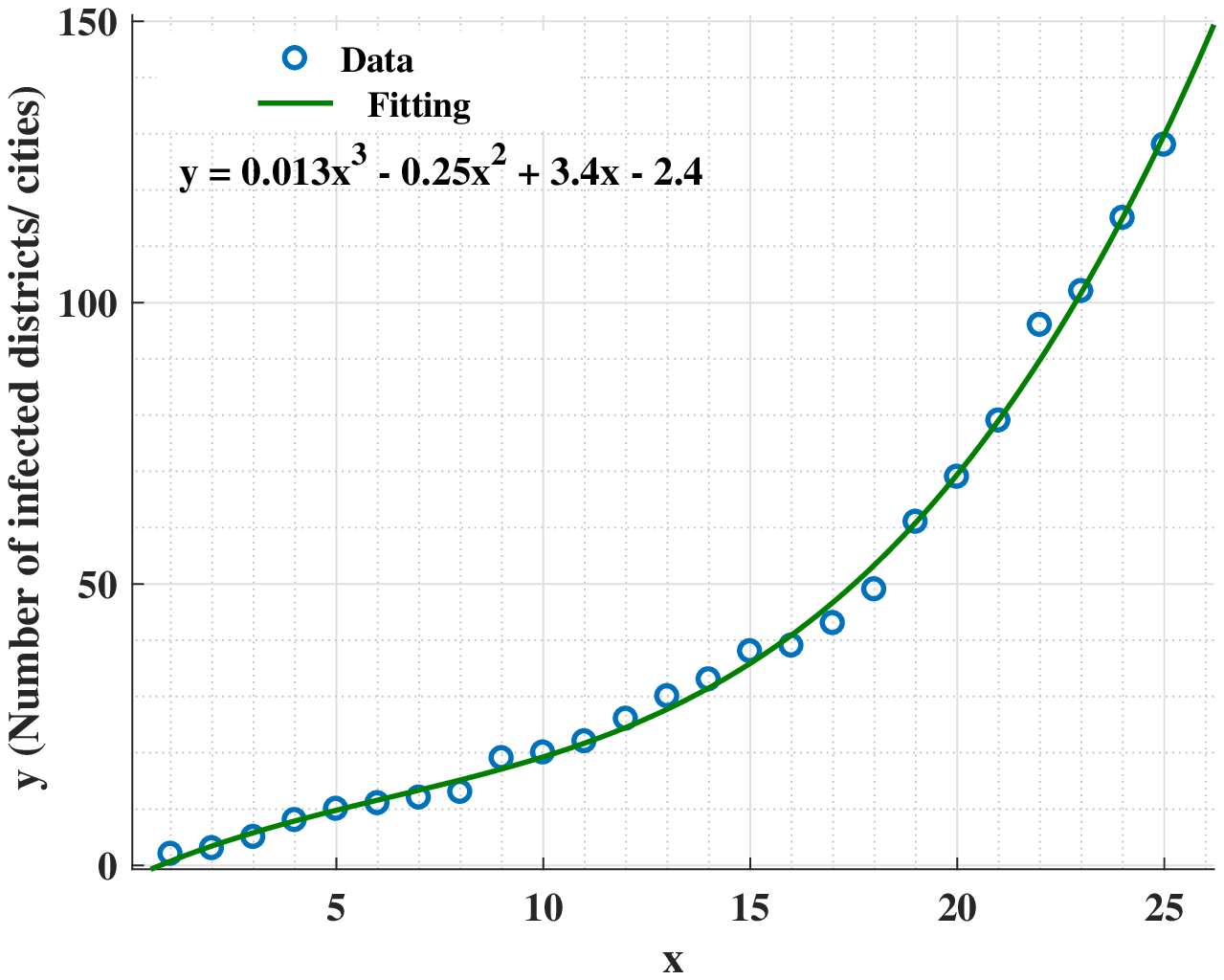}  
  \caption{}
  \label{fig17}
\end{subfigure}
\begin{subfigure}{.5\textwidth}
  \centering
 \includegraphics[width=1\linewidth]{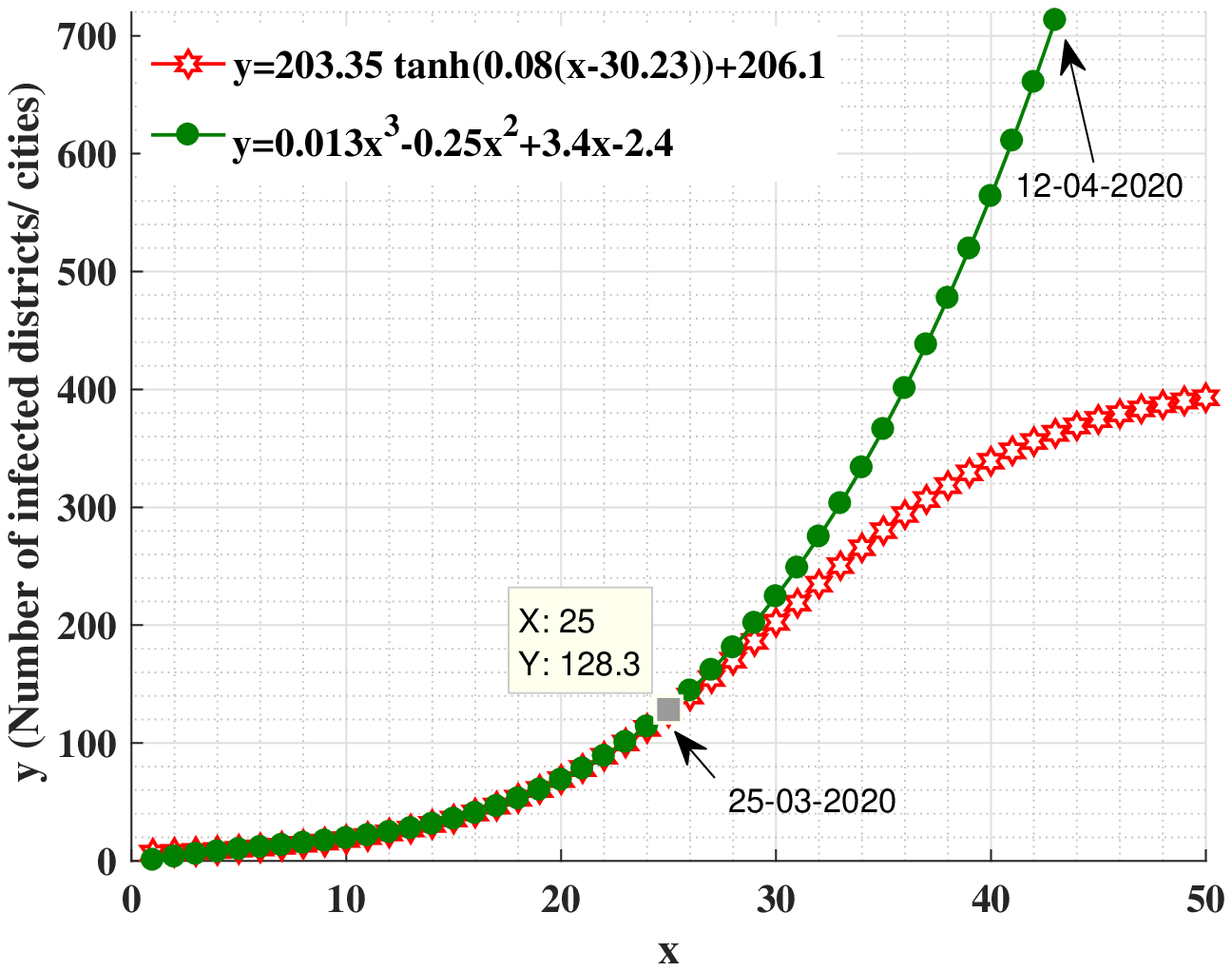}  
  \caption{}
  \label{fig18}
\end{subfigure}
\caption{(a) Approximate polynomial for the growth of cumulative number of districts or cities infected by COVID-19 before lockdown (b) The growth of number of infected districts or cities which follows the growth function before lockdown (in green), and the growth of number of infected districts or cities according to trend followed by the real data till April 17, 2020 (in red)}
\label{fig:fig1}
\end{figure}

The growth of $|V(t)|,$ the number of vertices in $V(t),$ as $t$ increases starting from January 30, 2020 to April 17, 2020 is plotted in Fig \ref{fig1}. This curve can be interpreted in many ways. First observe that it can be approximated by $\alpha \tanh (\beta (x+c)) + \gamma$ for some real values of $\alpha, \beta, \gamma, c$ (note that here $t$ considers positive integer values with a step size $1$, and $t=0$ on an initial date) when the day of lockdown acts as the origin. By data interpolation techniques, we show that the following hyperbolic function 
\begin{equation}\label{eqn:tan}
f(t)=203.35 \, \tanh(0.08 (t-30.23)) + 206.1
\end{equation} 
fits the data, as follows from Fig. \ref{fig17}. Here, in Fig \ref{fig17} the parameter $x$ in the $X$-axis counts the days starting from March 01, 2020 up to April 17, 2020.

It is obvious from Fig.~\ref{fig1} that the phenomena of lockdown makes a strong influence on the cumulative growth of the number of infected districts or cities. A sharp change of the slope of the curve can be observed just approximately two weeks after the lockdown i.e. March 25, 2020. We approximate the growth of the number districts infected with COVID-19 before lockdown. We find that the cumulative number of infected districts or cities is a cubic polynomial of the number of days starting from March 01, 2020. To be specific, the polynomial $$0.013\, x^3 - 0.25\, x^2 + 3.4\, x - 2.4$$ fits the growth, see Fig. \ref{fig17}. This polynomial is further extended to the entire domain to measure the impact of lockdown on $|V(t)|.$ Then we find that there could have been $713$ number of districts infected by cronavirus, which is close to the total number of districts by the April 12, 2020 compare to the $385$ many districts or cities on April 17, 2020, see Fig. \ref{fig18}. However, this should be treated as an observation but not as an absolute truth, since there are many other hidden parameters that can have significant impact on the growth of these phenomena.

It may be tempting to predict the number of districts or cities which will be infected with coronavirus in near future using the equation (\ref{eqn:tan}). However, this need not be done always. A noise has to be incorporated into the equation that can measure the uncertainties which depend on many parameters including the present medical facilities, testing statistics, and certain increase/decrease of communication across different districts. A model with significant number of parameters can reflect such growth, which we plan to explore in future.




\begin{figure}[ht]
\begin{subfigure}{.5\textwidth}
  \centering
 \includegraphics[width=1\linewidth]{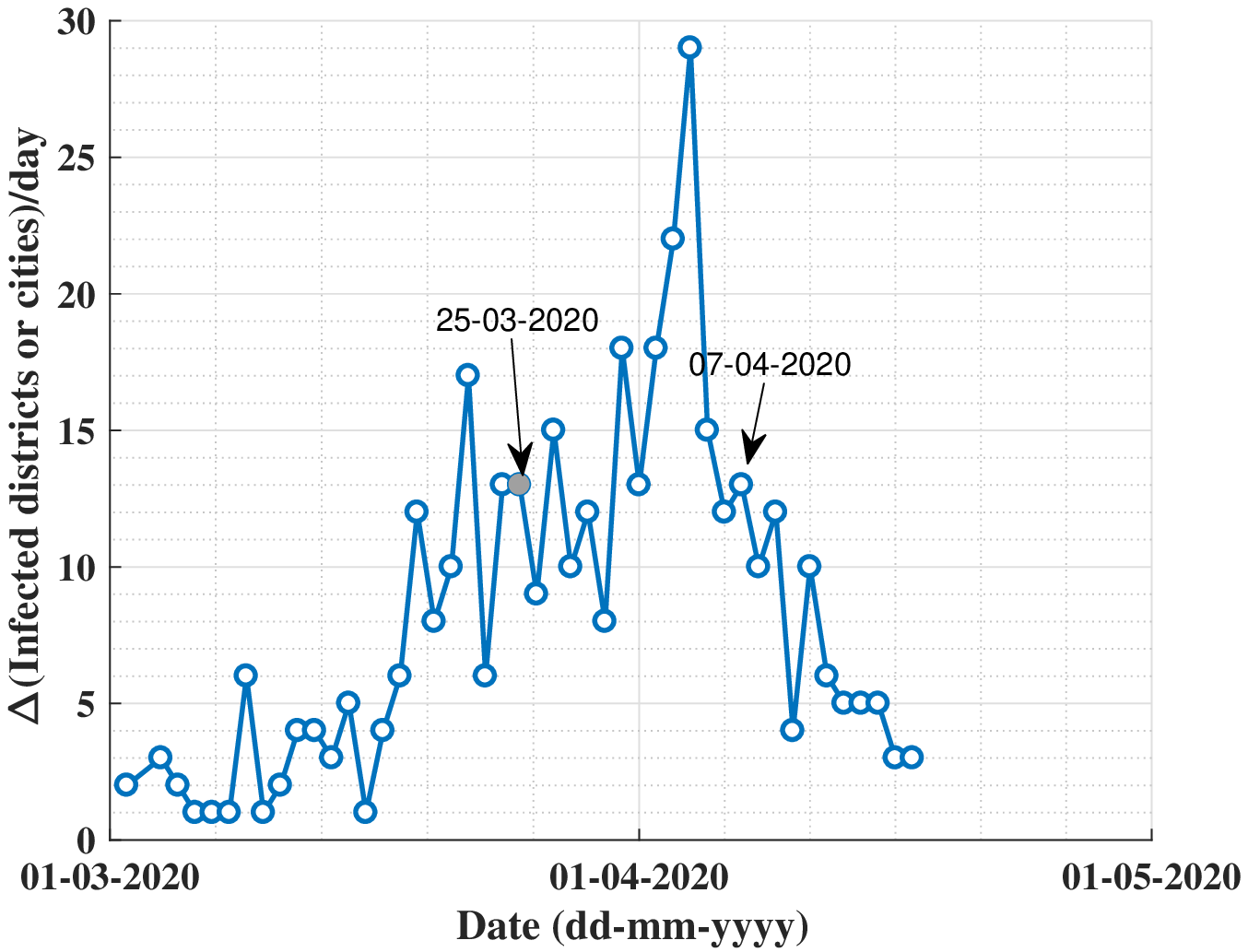}  
  \caption{}
  \label{fig15}
\end{subfigure}
\begin{subfigure}{.5\textwidth}
  \centering
 \includegraphics[width=1\linewidth]{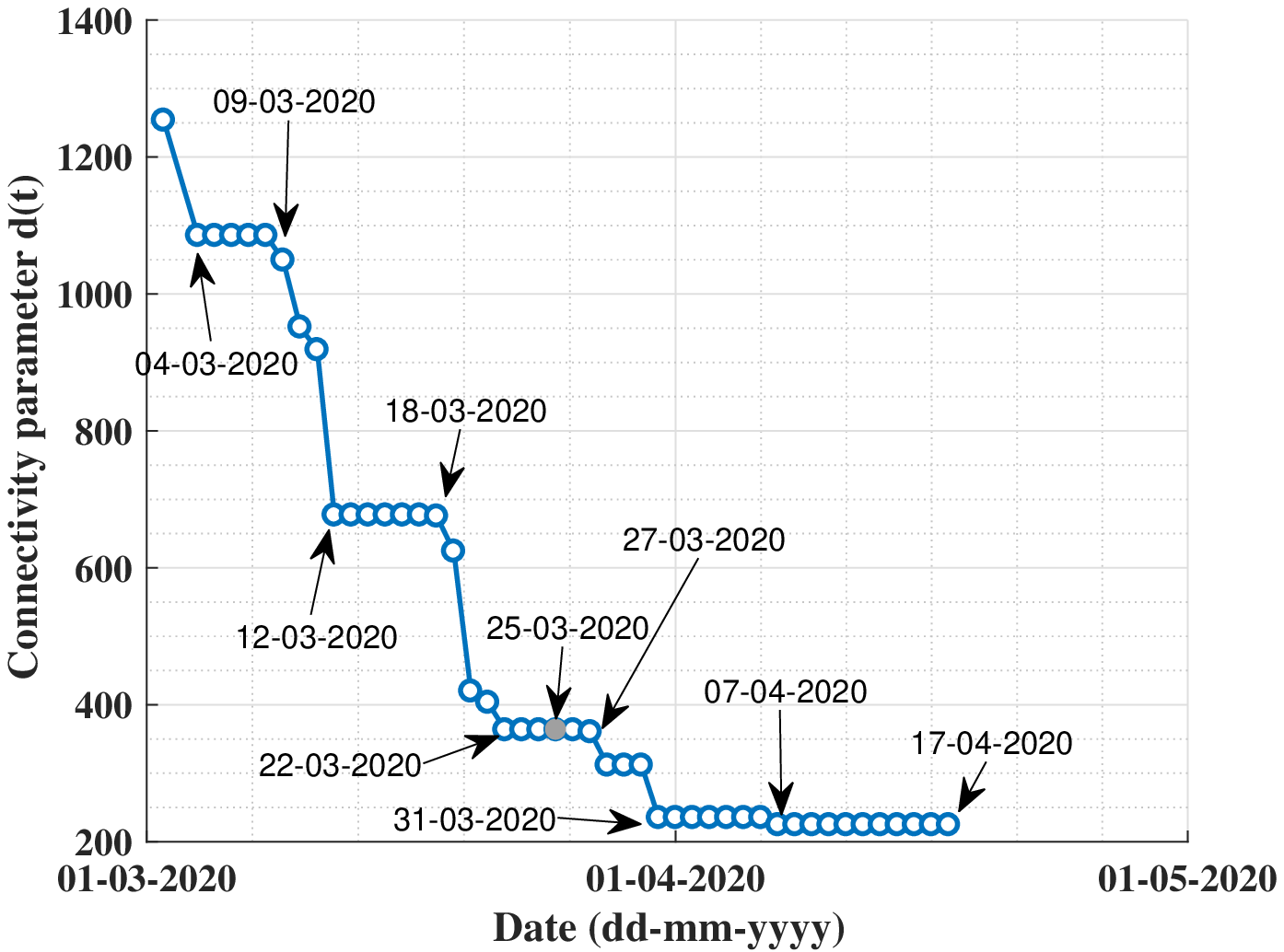}  
  \caption{}
  \label{fig2}
\end{subfigure}
\caption{(a) Number of new infected districts or cities on each date. (b) Value of the connectivity parameter of $N(t)$}
\label{fig:figr}
\end{figure}

In Fig. \ref{fig2}, the connectivity parameter $d(t)$ of the time series network $N(t)$ is plotted. In this plot, a strong effect of lockdown can be observed. Indeed, as people with COVID-19 have traveled through different regions of India without being detected, the distances between the infected districts are therefore large. In fact, during the first days after January 30, 2020, the next cases were reported in the main cities of India, notably Delhi, Hyderabad, Agra, Pune, Bangalore, Ladakh and Kancheepuram which are distant from any of the other. As the days go by, the distances between the infected districts or cities decrease, and finally after two weeks of lockdown, the distance stabilizes. The lockdown helped decrease the spread of the virus over long distances, but it did not stop the spread as follows from Fig. \ref{fig15}.

It has already been reported that there is a huge societal and economic impact of complete lockdown at the national level \cite {ecotimes1, ndtv1}. Therefore, lockdown can have a big impact on the lives of many people. It is very important to test almost everyone for COVID-19 before lifting the lockdown, otherwise the similar pattern of virus spread will be repeated. Another model to reduce the spread would be to lift the lockdown only in districts where no cases are reported, and people may be allowed to visit those districts when they are tested negative for the infection. Simultaneously, people living in districts with a high number of COVID-19 cases should be tested as soon as possible, and people who are not infected should be transported to other places where there is no one infected with COVID-19.



 \subsection{Spectral analysis}
 
 In this section we investigate the statistics of spectral radius (SR) and algebraic connectivity (AC) of $N(t)$, denoted by $\rho(t)$ and $\lam(t)$ respectively. It is needless to repeat the importance of these parameters in network analysis. In particular, it play an important role for the analysis of diffusion phenomena on networks \cite{spectra}. 
 
 \begin{figure}[ht]
\begin{subfigure}{.5\textwidth}
  \centering
  \includegraphics[width=1\linewidth]{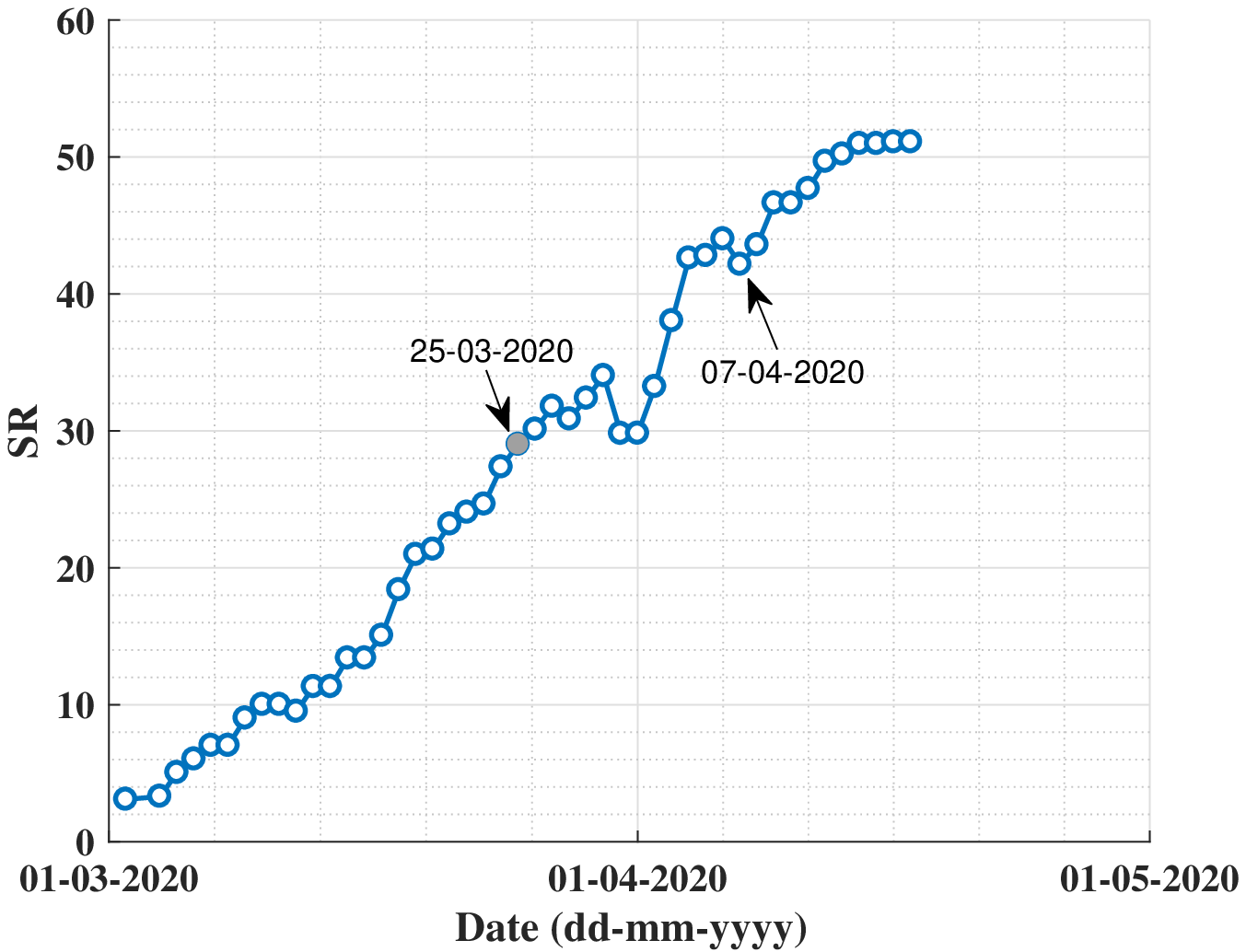}  
  \caption{}
  \label{fig3}
\end{subfigure}
\begin{subfigure}{.5\textwidth}
  \centering
  \includegraphics[width=1\linewidth]{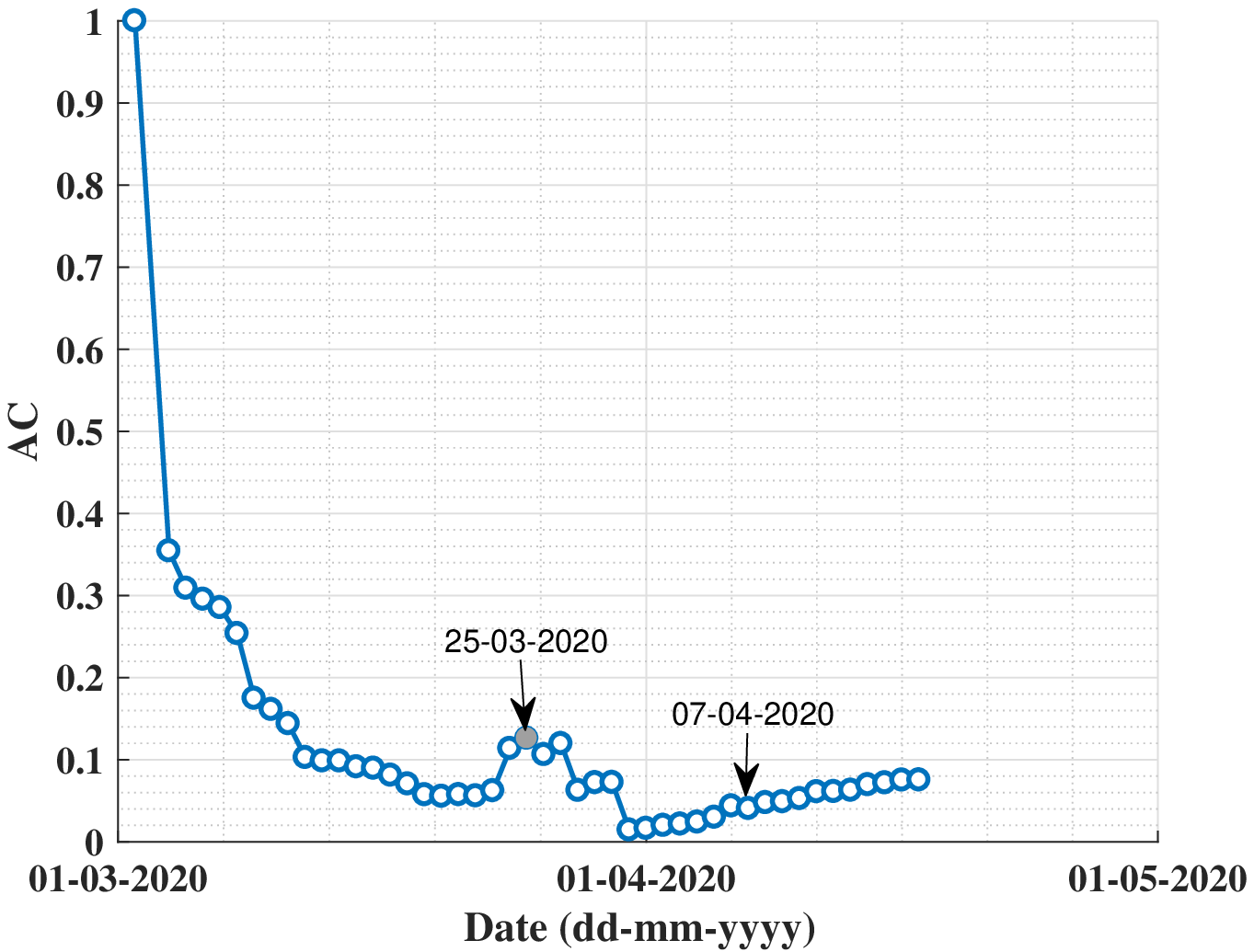}  
  \caption{}
  \label{fig4}
\end{subfigure}
\caption{(a) Spectral Radius (SR) and (b) Algebraic Connectivity (AC) of $N(t)$}
\label{fig:srac}
\end{figure}
 
Spectral radius of the network $N(t)$ is the largest absolute value of the eigenvalues of $A(t).$ It measures the robustness of a network against the spread of a virus on a network under the Susceptible-Infected-Susceptible (SIS) infection model \cite{Jamakovic06}. It is also well known that smaller the spectral radius, the higher the robustness of a network against the spread of viruses. $\rho(t)$ is plotted in Fig \ref{fig3}. Observe that $\rho (t) $ continues to increase except at one day. However, significant effect of lockdown is observed after almost two weeks of the lockdown and then the growth stabilizes. This means that if a SIS model is executed on these networks then the risk of the nodes getting affected increases and it stabilizes after two weeks of lockdown when the growth of $\rho(t)$ is negligible. 
 
On the other hand, $\lambda (t),$ measures the network connectivity, a smaller value of $\lam (t)$ implies that the network $N (t)$ tends to collapse into components connected \cite {Jamakovic08}. It also acts as a deterrent for diffusion processes on networks  \cite{ac} \cite{ac1}. $\lam (t)$ of $N(t)$ is plotted in Fig \ref {fig4}. It follows that $\lam (t)$ gradually decreases until lockdown except on a few dates, which means that the propagation of the virus occurred mainly in remote places, but after nearly two weeks of lockdown, $\lam (t)$ is a strictly increasing function. This means that the virus spreads evenly across districts after April 1, 2020, after a week of lockdown.


\subsection{Influential spreaders}

\begin{figure}[ht]
\begin{subfigure}{.5\textwidth}
  \centering
  \includegraphics[width=1\linewidth]{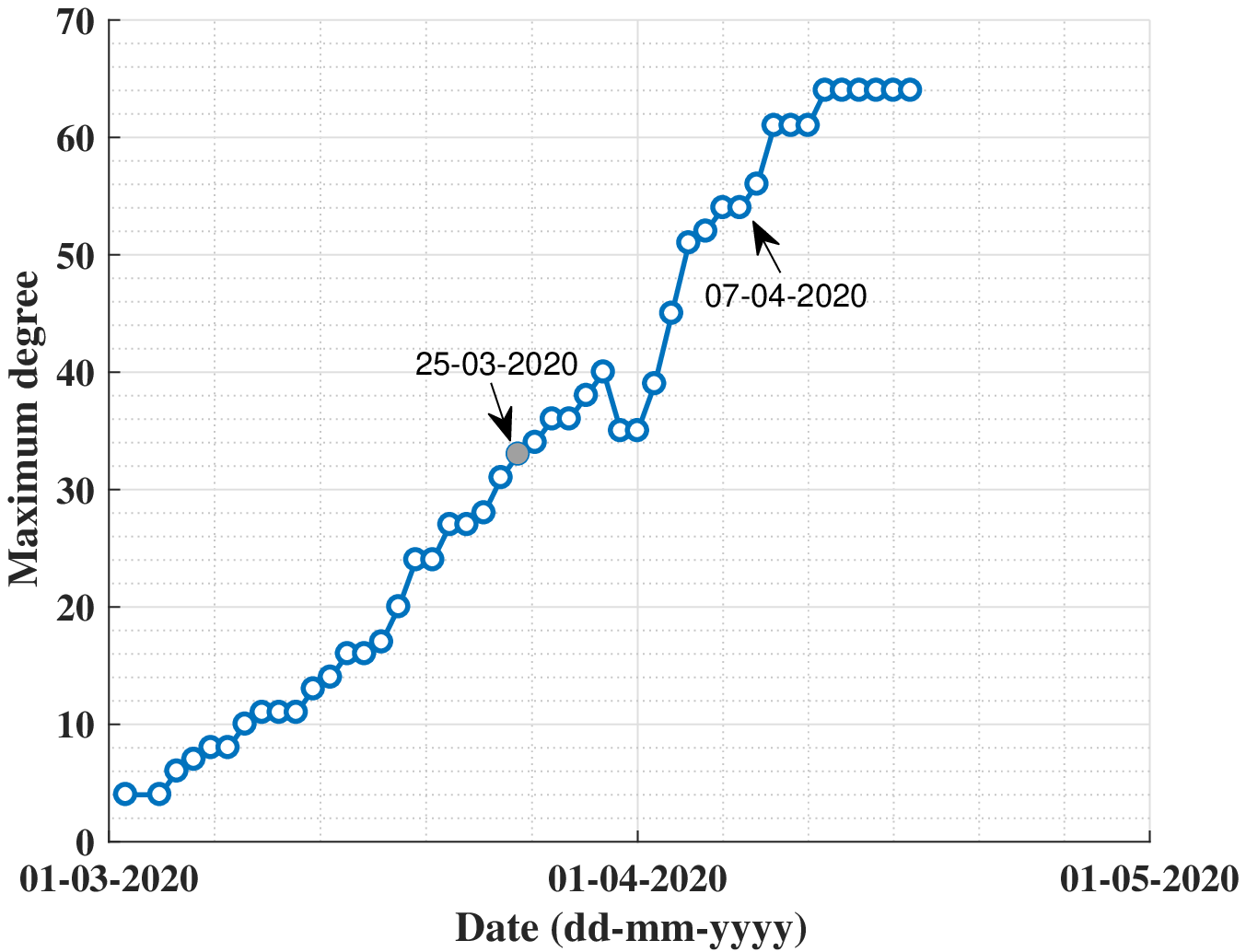}  
  \caption{}
  \label{fig5}
\end{subfigure}
\begin{subfigure}{.5\textwidth}
  \centering
  \includegraphics[width=1\linewidth]{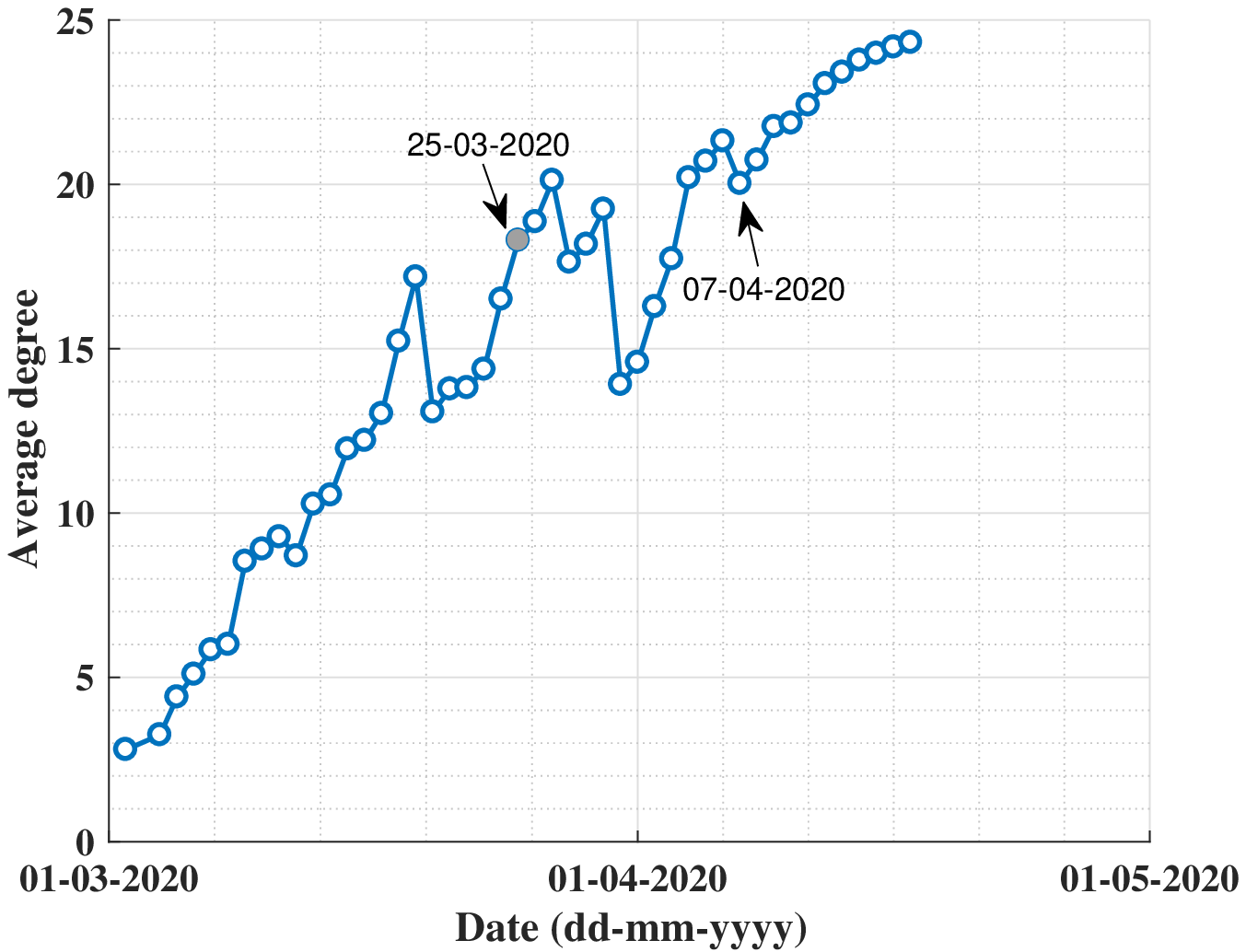}  
  \caption{}
  \label{fig6}
\end{subfigure}
\caption{(a) Maximum degree, and (b) average degree of networks obtained from the data.}
\label{fig:fig4}
\end{figure}

The degree of a vertex in a network is the number of other vertices it is connected with edges. The maximum degree of a network is maximum value of the degrees of vertices of the network.  In the context of diffusion on networks the maximum degree vertices are known to be influential spreaders. Thus an increment in maximum degree of a time series network data implies that newly appeared vertices get connected with existing high degree vertices of the networks. The maximum degree of $N(t)$ is plotted in Fig \ref{fig5}, and it can be seen that it increases sharply until almost two weeks after lockdown. Then the maximum degree stabilizes and hence spreading of the virus from high degree vertices is stopped.\footnote{We exclude the list of influential spreaders (districts) in this note.}

The average degree of a network $N$ on $n$ vertices is defined as $$\frac{1}{n}\sum_{i} \mbox{deg}(i)$$ where $\mbox{deg}(i)$ denotes the degree of the vertex $i.$ 
The average degree of a network signifies densification of the network. In the context of diffusion of information increment of average degree of a time series network data implies local diffusion \cite{ac2}. It can be observed from the Fig \ref{fig6} that it increases with time $t$ for the network $N(t)$ throughout except at a few dates, and after almost two weeks of lockdown it  increases strongly. It is therefore an indication of the spread of the virus in neighboring districts  


\subsection{Clustering analysis}

The clustering coefficient (cc) of a vertex $v$ of a network $N $ with the set of edges $E,$ is defined as follows $$cc(v) = \frac{|\{e_{uw} : u, w\in N_v, e_{uw}\in E \}|}{{\mbox{deg}(v) \choose 2}}$$ where $e_{xy} $ means an edge which connects the vertices $x$ and $y,$ and $N_v$ denotes the set of vertices which are linked to $v$ by an edge. A triangle in a graph is a sub-graph on three vertices so that each pair of vertices is linked by an edge. Thus $cc(v)$ can be defined as the ratio of the number of triangles in $N$ attached to $v$ to the number of all possible triangles attached to the neighbors of $v.$ The average clustering coefficient of the network $N$ on the nodes $ n $ is defined as $$CC = \frac{\sum_v cc(v)}{n}$$ where the summation extends over all the vertices of $N.$ Thus, CC of a network measures to what extend the edges are clustered in triangles on average. A high CC value of a network indicates that most of the edges of the network are attached to triangles. On the other hand, a smaller CC value means that there are more number of edges in the network that are not part of triangles than the number of edges which are part of triangles.

\begin{figure}[ht]
\begin{subfigure}{.5\textwidth}
  \centering
  \includegraphics[width=1\linewidth]{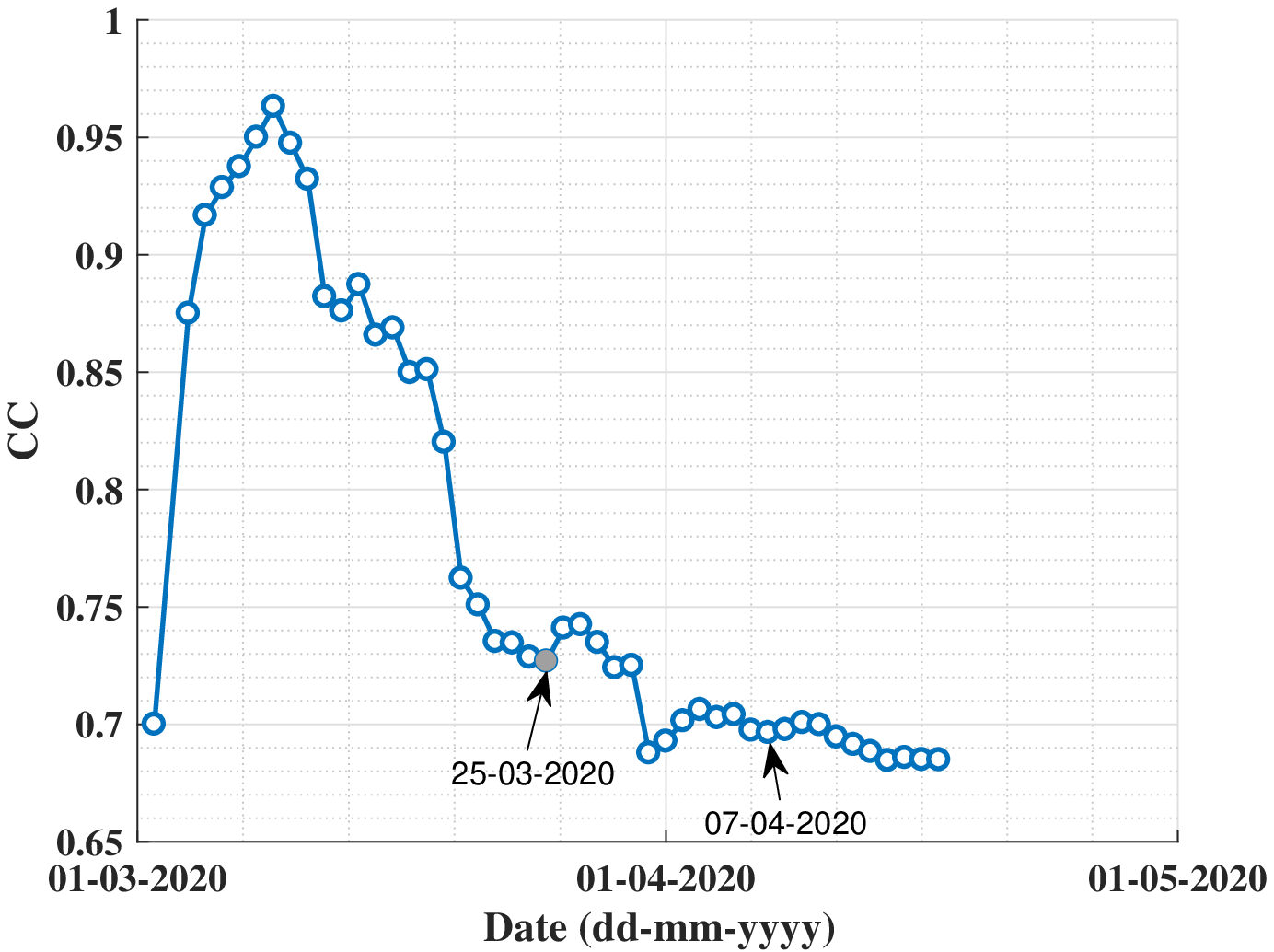}  
  \caption{}
  \label{fig7}
\end{subfigure}
\begin{subfigure}{.5\textwidth}
  \centering
  \includegraphics[width=1\linewidth]{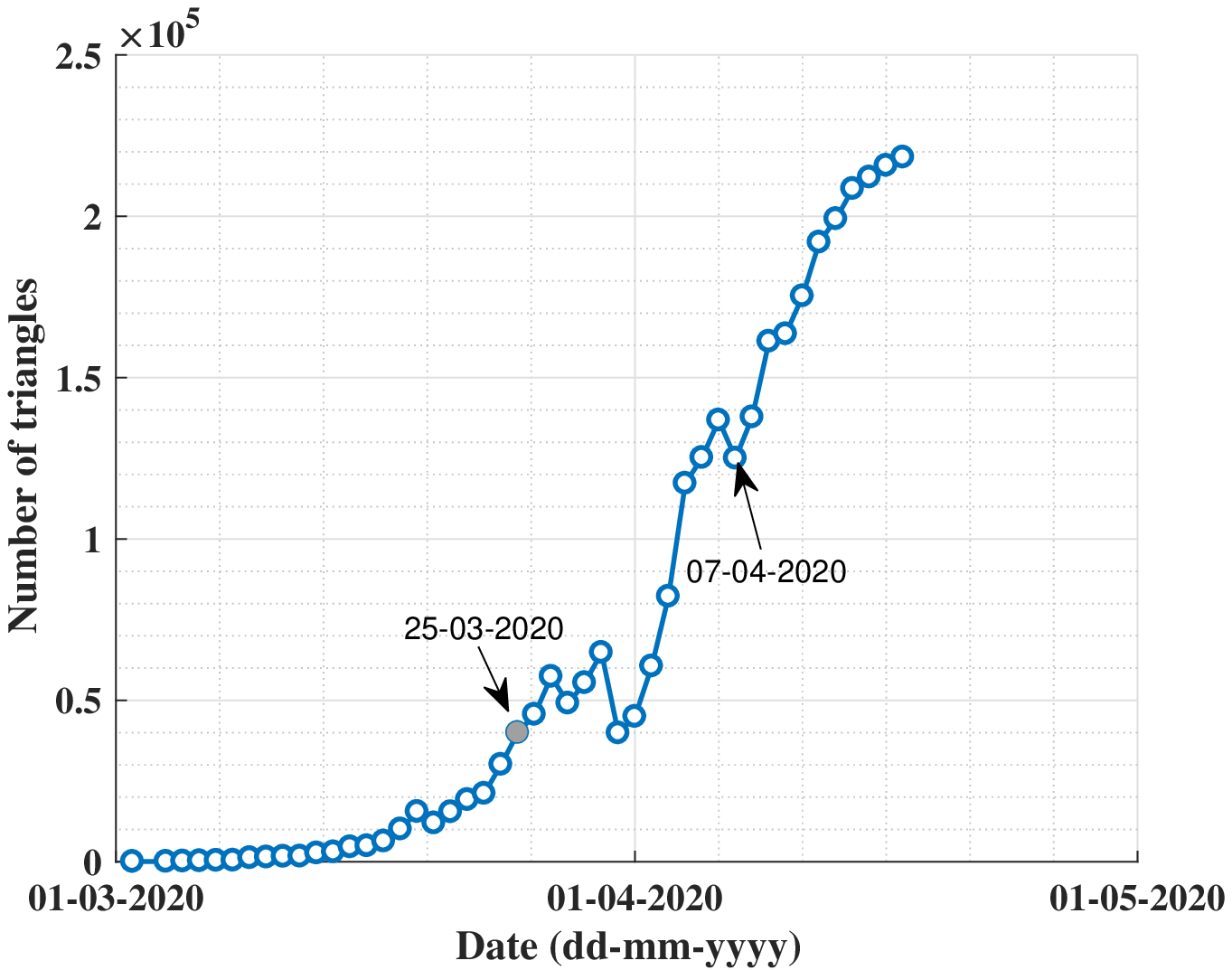}  
  \caption{}
  \label{fig8}
\end{subfigure}
\caption{(a) Clustering coefficient, and (b) number of triangles in $N(t)$}
\label{fig:fig}
\end{figure}

In Fig.  \ref {fig7}, the average clustering coefficients of $ N (t)$ are plotted. We can see that CC had a very high value (almost $1$) before the lockdown compared to its value after the lockdown. In addition, after almost two weeks of lockdown, the CC value almost stabilizes with the increase in the number of vertices. This means that the number of triangles is significantly less than the number of newly appeared edges in $N(t)$ as $t $ increases. There are therefore more legs in the network $N(t)$ as $t$ increases.

The triangles in a network signify a local clustering and a trapping of the diffusion \cite{ac3}. In Fig. \ref{fig8}, the number of triangles of $N (t)$ is plotted. This shows that the number of triangles increases sharply with the addition of newly infected districts after two weeks of lockdown.

In general, it is an established phenomenon that the higher the CC of a network, the higher the number of triangles. However, in this case, note that the number of triangles increases but this does not affect the CC value of the network. One of the possible reasons for that is the rate of change in terms of the number of triangles is same as rate of change of $deg(i)(deg(i)-1)/2$ for all the vertices $i$ of the network.


\subsection{Shrinking diameter phenomena and average path length}

\begin{figure}[ht]
\begin{subfigure}{.5\textwidth}
  \centering
  \includegraphics[width=1\linewidth]{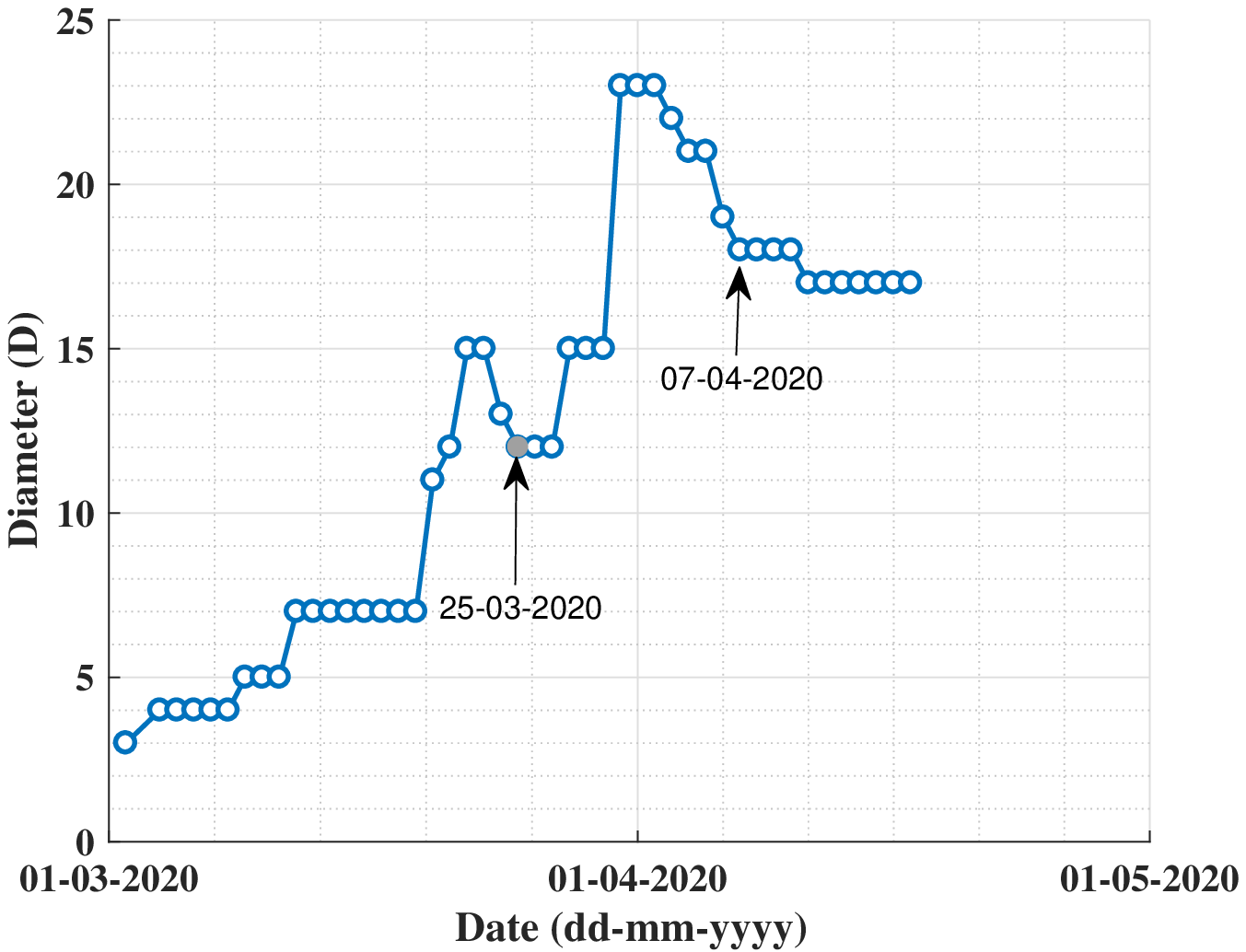}  
  \caption{}
  \label{fig9}
\end{subfigure}
\begin{subfigure}{.5\textwidth}
  \centering
  \includegraphics[width=1\linewidth]{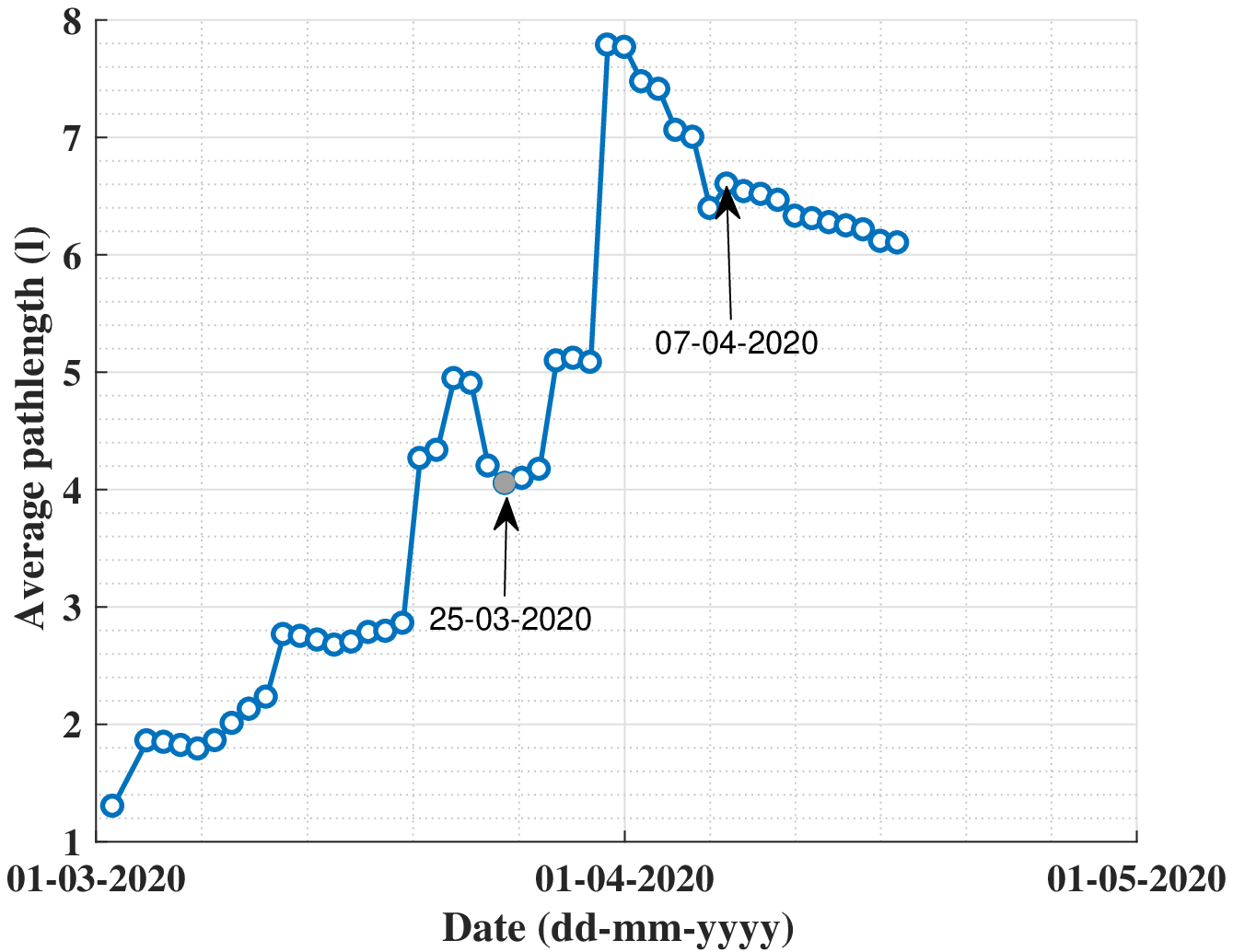}  
  \caption{}
  \label{fig10}
\end{subfigure}
\caption{Statistics of (a) diameter, and (b) average path length of $N(t)$}
\label{fig:fig}
\end{figure}

The diameter of a network is the longest of the shortest distances between all the pairs of nodes in the network. The diameter of $N (t)$ is plotted in Fig \ref{fig9}. We can observe that before lockdown, the diameter gradually increases except on a few dates, then a week after lockdown, it began to decrease. This is due to the fact that the initial spread of the virus was dispersed to remote locations all over India. Later, local growth at different intermediate places causes the diameter to drop.

The average path length of a network $N$ on $n$ nodes is defined as $$l=\frac{\sum_{u\neq v} d(u, v)}{n(n-1)}$$ where $d (u, v)$ denotes the shortest number of edges to reach from $u$ to $v$ on $N$ and vice-versa.  A smaller value of $l$ of a network means a rapid traveling between any pair of vertices via the edges of the network. We can see in Fig \ref{fig10} that the average path length of $N(t)$ increases with $t$ which happened due to the random appearance of different vertices in distant places. However, after about two weeks of lockdown, the value of $l$ decreases, reflecting local growth across the network.

 

\subsection{Community analysis}

Communities are sub-graphs of a network such that the density of edges within a community is comparatively higher than the density of edges between two separate communities \cite{ac4}. The community structure has a huge impact on the phenomena of diffusion on networks \cite{ac5}. Modularity, denoted $ Q $, is a quantitative measure for the detection of networked communities. Higher value of $Q$ implies that the communities are well separated. The size of a community is the number of vertices within the community.

\begin{figure}[ht]
\begin{subfigure}{.5\textwidth}
  \centering
  \includegraphics[width=1\linewidth]{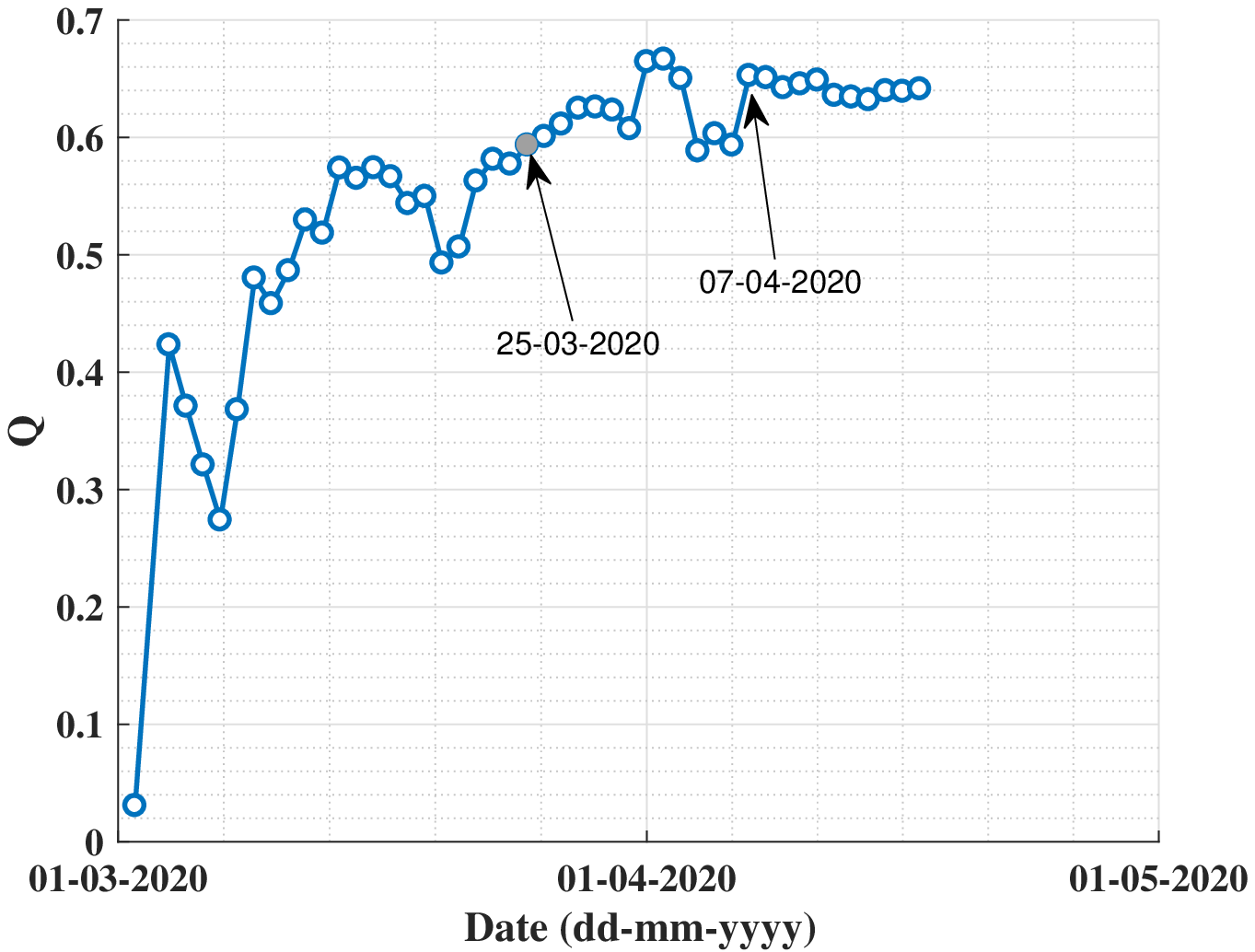}  
  \caption{}
  \label{fig11}
\end{subfigure}
\begin{subfigure}{.5\textwidth}
  \centering
  \includegraphics[width=1\linewidth]{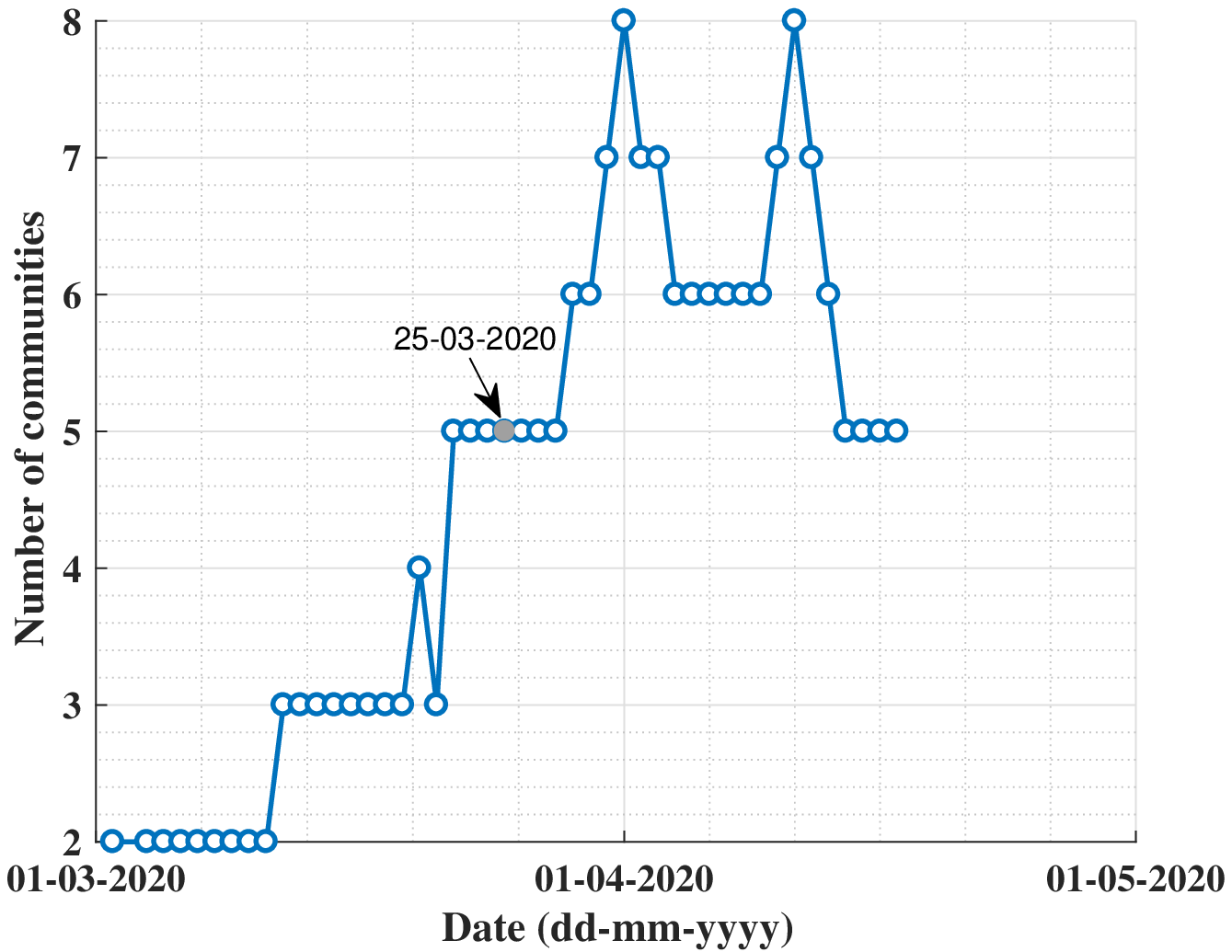}  
  \caption{}
  \label{fig12}
\end{subfigure}
\caption{(a) Modularity value $Q$ for community detection of $N(t)$ \, (b) The number of communities in $N(t)$}
\label{fig:fig}
\end{figure}

\begin{figure}[ht]
  \centering
\includegraphics[width=0.6\linewidth]{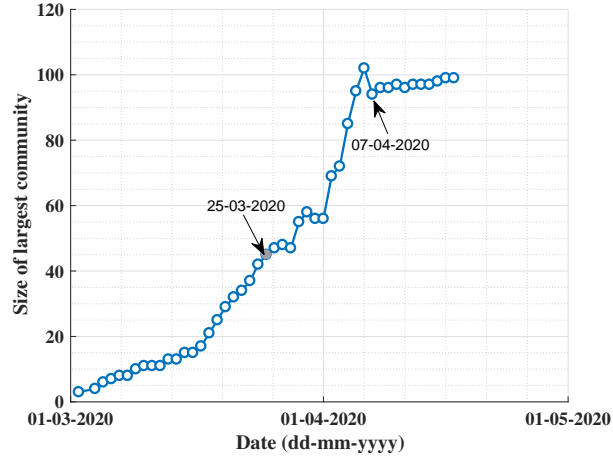}  
\caption{Growth of size of the largest community in $N(t)$}
\label{fig13}
\end{figure}

The $Q$ value, number of communities and maximum size of the communities of $N(t)$ are plotted in Fig \ref{fig11}, Fig \ref{fig12}, and Fig \ref{fig13} respectively. The communities of $N(t)$ are determined using the well-known Louvain method \cite{ac6} with higher modularity values. Observe that the number of communities increases before the lockdown on March 25, 2020. Then almost two weeks after the lockdown, the number of communities stabilizes with the exception of the peaks on a few dates. Again, as mentioned in the case of other parameters, this happened due to the local growth in the number of districts or cities after the lockdown. It can also be noted that the size of the largest community increased considerably before the lockdown, and stabilized just after almost two weeks of the lockdown. This extends the spread of the virus to neighboring districts or cities. The largest community includes approximately 100 of districts or cities out of a total of nearly 430 of districts or cities affected by SARS-CoV-2.


\section{Conclusion}

We analyzed time series data of number of districts or cities in India that were infected with COVID-19 from March 1 to April 17, 2020. We studied several statistics from this data as part of the network data. These networks are defined using the geodesic distances between the districts.

It is established by the analysis that lockdown helped preventing the spread of the virus over large distance, but could not stop the spread across neighboring districts. The lockdown has effected strongly on the spread after about two weeks. 

Moreover, the following speculations can be made based on the analysis of the data performed in this note. \begin{itemize}
    \item[(a)] Sudden lifting of lockdown may not be an intelligent idea, as it may speed up the spread of  SARS-CoV-2 over large distances, which is now controlled due to lockdown.
    
    \item[(b)] The lockdown can be lifted in places where no one is infected with the virus, and medical and testing facilities should be developed around these areas. Uninfected people can only be allowed to visit these places.
    
    \item[(c)] Immediate test facilities should be constructed at the overlap or at the boundaries of the communities identified in this note. Note that priority should be given to the neighboring districts of these communities in order to stop the spread to neighboring districts, which could possibly prevent the merging of communities.

    \item[(d)] Evacuation plans must be ready to move healthy people from the heavily infected areas.
    
    \item[(e)] Methods need to be developed for mass testing of COVID-19.
    
\end{itemize}

Finally, this study suggests that the lockdown may not save us from the spread of SARS-CoV-2, but a sincere, disciplined human race can.\\\\

\noindent{\bf Acknowledgment.} 
The authors thank Vaidik Dalal of How India Lives for his help with the data. Thanks are also due to Buddhananda Banerjee for his helpful comments and suggestions on the topic of this note.

\end{document}